\documentclass[a4paper,fleqn,usenatbib]{mnras}

\usepackage{newtxtext,newtxmath}

\usepackage[T1]{fontenc}
\usepackage{ae,aecompl}

\usepackage{graphicx}	
\usepackage{amsmath}	
\usepackage{amssymb}	

\def\msun{\hbox{${\rm M}_{\odot}$}}

\def\mspy{\hbox{${\rm M}_{\odot}$\,yr$^{-1}$}}
\def\rsun{\hbox{${\rm R}_{\odot}$}}

\def\mstar{\hbox{$M_{\star}$}}
\def\rstar{\hbox{$R_{\star}$}}

\def\teff{\hbox{$T_{\rm eff}$}}
\def\logg{\hbox{$\log g$}}
\def\sn{\hbox{S/N}}
\def\vrad{\hbox{$v_{\rm rad}$}}

\def\kms{\hbox{km\,s$^{-1}$}}
\def\vsini{\hbox{$v \sin i$}}

\def\BmV{\hbox{${\rm B-V}$}}
\def\VmRj{\hbox{${\rm V-R_{\rm J}}$}}

\def\AV{\hbox{$A_{\rm V}$}}

\def\degr{\hbox{$^\circ$}}

\newcommand{\caii}{\hbox{Ca$\;${\sc ii}}}

\newcommand{\hei}{\hbox{He$\;${\sc i}}}

\title[The magnetic propeller accretion regime of LkCa~15]{The magnetic propeller accretion regime of LkCa~15} 
\author[J.-F.~Donati et al.]{J.-F.~Donati$^{1}$\thanks{E-mail: jean-francois.donati@irap.omp.eu},
J.~Bouvier$^{2}$, S.H.~Alencar$^{3}$, C.~Hill$^1$, A.~Carmona$^1$, C.P.~Folsom$^1$, 
\newauthor F.~M\'enard$^{2}$, S.G.~Gregory$^{4}$, G.A.~Hussain$^5$, K.~Grankin$^6$, C.~Moutou$^7$, 
\newauthor L.~Malo$^8$, M.~Takami$^{9}$, G.J.~Herczeg$^{10}$ and the MaTYSSE collaboration 
\vspace{1mm}
\\ 
$^1$ Univ.\ de Toulouse, CNRS, IRAP, 14 avenue Belin, 31400 Toulouse, France \\ 
$^2$ Univ.\ Grenoble Alpes, CNRS, IPAG, 38000 Grenoble, France \\ 
$^3$ Departamento de F\`{\i}sica -- ICEx -- UFMG, Av. Ant\^onio Carlos, 6627, 30270-901 Belo Horizonte, MG, Brazil \\
$^4$ SUPA, School of Physics and Astronomy, Univ.\ of St~Andrews, St~Andrews, Scotland KY16 9SS, UK \\
$^5$ ESO, Karl-Schwarzschild-Str.\ 2, D-85748 Garching, Germany \\
$^6$ Crimean Astrophysical Observatory, Nauchny, Crimea 298409 \\
$^7$ CFHT Corporation, 65-1238 Mamalahoa Hwy, Kamuela, Hawaii 96743, USA \\
$^8$ D\'epartement de physique, Universit\'e de Montr\'eal, C.P.~6128, Succursale Centre-Ville, Montr\'eal, QC, Canada  H3C 3J7 \\
$^9$ Institute of Astronomy and Astrophysics, Academia Sinica, PO Box 23-141, 106, Taipei, Taiwan \\
$^{10}$ Kavli Institute for Astronomy and Astrophysics, Peking University, Yi He Yuan Lu 5, Haidian Qu, Beijing 100871, China 
}

\date{Submitted 2018 September -- Accepted 2018 xxx } 

\pubyear{2019}

\begin{document}

\label{firstpage}
\pagerange{\pageref{firstpage}--\pageref{lastpage}}
\maketitle

\begin{abstract}
We present a spectropolarimetric study of the classical T~Tauri star (cTTS) LkCa~15 investigating the large-scale magnetic 
topology of the central star and the way the field connects to the inner regions of the accretion disc.  We find that 
the star hosts a strong poloidal field with a mainly axisymmetric dipole component of 1.35~kG, whereas the 
mass accretion rate at the surface of the star is $10^{-9.2}$~\mspy.  It implies that the magnetic field of 
LkCa~15 is able to evacuate the central regions of the disc up to a distance of 0.07~au at which the Keplerian 
orbital period equals the stellar rotation period.  Our results suggest that LkCa~15, like the lower-mass cTTS AA~Tau, 
interacts with its disc in a propeller mode, a regime supposedly very efficient at slowing down the rotation of cTTSs 
hosting strong dipolar fields.  
\end{abstract}

\begin{keywords}
stars: magnetic fields --
stars: formation --
stars: imaging --
stars: individual:  LkCa~15  --
techniques: polarimetric
\end{keywords}



\section{Introduction}
\label{sec:int}

Although major progress was achieved in the last few decades in understanding how low-mass stars and their planets form, 
many critical phases in this evolution are still poorly understood and are waiting for a consistent physical explanation.  
The way protostars and their protoplanetary accretion discs succeed at expelling the initial angular momentum and 
magnetic flux inherited from the parent molecular cloud, before ending up as slowly rotating T~Tauri stars with much weaker
fields than flux conservation would imply, is one of them.  The last step of this process, where the newly formed low-mass stars 
(called classical T~Tauri stars / cTTSs when they still accrete from their discs) interact with the core disc regions  
through magnetic funnels and apparently succeed at locking their rotation on that of the inner disc, has 
been the subject of many theoretical and observational studies \citep[][]{Bouvier14}.  

To progress on this issue one critically needs observational constraints on the large-scale fields that control the 
star-disc interaction and the loss of angular momentum, and on how these fields relate to the observed diversity of 
accretion modes that continuous photometric campaigns from space probes like CoRoT or K2 revealed \citep{Cody14, Sousa16, Cody18}.  
The main mechanism identified to date as capable of efficiently slowing down the rotation of cTTSs is the magnetic propeller 
or magnetospheric ejection model 
\citep[e.g.,][]{Romanova04, Ustyugova06, Zanni13}, in which the magnetic field of the central star is strong enough to 
truncate the inner accretion disc at or beyond the corotation radius and efficiently expel angular momentum outwards.  

For stellar parameters typical to cTTSs (e.g., 1~\msun, 2~\rsun\ and a rotation period of 8~d), the dipole component of the 
large-scale field needs to be stronger than 1--4~kG for accretion rates in the range $10^{-9}$ to $10^{-8}$~\mspy\ \citep{Bessolaz08}.  
However, very few cTTSs are known to host magnetic fields with such strong dipole components;  only the prototypical cTTS AA~Tau 
was unambiguously found to be in such a state \citep{Donati10}, whereas others like V2129~Oph \citep{Donati11} may only sporadically reach 
it.  Hence the need to explore the large-scale fields of a wide sample of cTTSs and unveil the strengths of the dipole component of their 
fields, to confirm whether the magnetic propeller is indeed the main process that forces cTTSs into slow rotation.  

This can be achieved using phase-resolved spectropolarimetric observations coupled to tomographic imaging techniques inspired 
from medical imaging \citep[e.g.,][]{Donati06b, Donati09}.  In this paper, we study the well known cTTS LkCa~15, whose accretion 
disc, called a transition disc from the fact that it features a wide dust gap from the inner disc up to 50~au from the central star, 
is reported to be potentially warped \citep{Oh16} with claims of ongoing planet formation \citep[][later challenged by 
\citealt{Thalmann15,Thalmann16, Mendigutia18}]{Kraus12b, Sallum15}.  Whereas our paper concentrates on the magnetic field of LkCa~15, a companion 
study focusses on the properties of its inner accretion disc \citep{Alencar18}.  This companion paper demonstrates in particular that, 
like AA~Tau \citep{Bouvier07b, Esau14}, LkCa~15 is a periodic 'dipper' regularly eclipsed by a dusty inner disc warp connected to accretion 
funnels and crossing the line of sight as the star rotates, with line profile variations and veiling variability consistent with a 
highly inclined inner disk interacting with the stellar magnetosphere.  Following a short review of the evolutionary status of LkCa~15, 
we carry out our tomographic imaging study and discuss its implications for our understanding of stellar formation.

\section{Evolutionary status of LkCa~15}
\label{sec:evo}

To ensure homogeneity and consistency with previous MaPP and MaTYSSE papers, we start our study with a short revision 
of the evolutionary status of LkCa~15, incorporating the latest relevant measurements from the literature.  

Applying our spectral classification tool \citep[][]{Donati12} to our best data (see Sec.~\ref{sec:obs}), 
we obtain that the photospheric temperature and logarithmic surface gravity are respectively equal to 
$\teff=4500\pm50$~K and $\logg=4.0\pm0.1$ (in cgs units), in good agreement with results of \citet{Alencar18}.   
From its measured \BmV\ photometric color of 1.26 \citep[][]{Grankin08} and the one we can expect of a young star 
of this temperature \citep[$1.10\pm0.02$,][]{Pecaut13}, we obtain a first estimate for the visual extinction of 
$\AV=0.68\pm0.20$;  averaging with the literature value derived from the \VmRj\ color \citep[$\AV=0.41$,][]{Grankin08}, 
we adopt a value of $\AV=0.55\pm0.20$.  We also obtain a bolometric correction of $-0.64\pm0.02$ \citep{Pecaut13}.  

Starting from a maximum V magnitude of $11.87\pm0.10$ \citep{Grankin08} and using the latest distance estimate 
from Gaia \citep[$158.8\pm1.3$~pc,][]{Gaia18}, we obtain for LkCa~15 a bolometric magnitude of $4.68\pm0.25$, i.e., 
a logarithmic luminosity relative to the Sun of $0.02\pm0.10$.  Comparing with the PMS evolutionary models of 
\citet[][assuming solar metallicity and including convective overshooting]{Siess00}, we find that LkCa~15 is 
a $\mstar=1.25\pm0.10$~\msun\ star with a radius of $\rstar=1.6\pm0.2$~\rsun\ and an age of 
$\simeq$5~Myr.  
These models further indicate that LkCa~15 is no longer fully- but presumably still largely convective, with a convective 
depth of $\simeq$0.55~\rstar, and is expected to turn largely radiative (in radius) within less than a few Myr.  
Compared to our previously studied cTTSs, we note that LkCa~15 is similar to, though slightly less massive and older 
than, the well-known cTTS V2129~Oph \citep{Donati11}.  

Given the rotation period of LkCa~15 \citep[$5.70\pm0.10$~d,][see also Sec.~\ref{sec:obs}]{Alencar18} and its line-of-sight 
projected rotation velocity ($\vsini$$\simeq$13~\kms, see Sec.~\ref{sec:tom}), we obtain that $\rstar \sin i=1.46\pm0.10$~\rsun, 
and thus that $i$, the angle between the rotation axis and the line of sight, ranges between 50\degr\ and 90\degr.  
Whereas this remains compatible with the inclination angle of the outer disc \citep[$\simeq$50\degr,][]{Thalmann14, vanderMarel15}, 
our tomographic modeling (see Sec.~\ref{sec:tom}) suggests that $i$ needs to be at least 70\degr\ to ensure an optimal 
fit to the spectropolarimetric data, implying that the inner disc is likely warped as already suggested by previous 
studies \citep{Oh16}.  This is also supported by the fact that LkCa~15 is a periodic 'dipper' \citep{Alencar18}, with the 
inner disc warp regularly occulting the star as it rotates, and requiring $i$ to be within $70-90$\degr\ \citep{Cody18}.

\section{Spectropolarimetric observations}
\label{sec:obs}

Our set of observations, carried within the MaTYSSE programme \citep{Donati14}, consists of 14 circularly polarized spectra 
collected in late 2015 with the ESPaDOnS 
spectropolarimeter at Canada-France-Hawaii Telescope (CFHT), covering 370 to 1,000~nm at a resolving power of
65,000 \citep{Donati03}.  Raw frames were reduced with the standard ESPaDOnS reduction package, and Least-Squares 
Deconvolution \citep[LSD,][]{Donati97b} was applied to all spectra, using a line list appropriate to LkCa~15.  
The full journal of observations is presented in Table~\ref{tab:log}.  

\begin{table}
\caption[]{Journal of ESPaDOnS observations of LkCa~15.  All observations consist of sequences of 4 subexposures, 
each lasting 765~s.  Columns respectively list, for each observation, the UT date, time, Barycentric Julian Date 
(BJD), peak signal to noise ratio \sn\ (per 2.6~\kms\ velocity bin), rms noise level in Stokes $V$ LSD profiles, 
and rotation cycle $r$ computed using ephemeris BJD (d)~=~$2457343.8+5.70 r$ as in \citet{Alencar18}. }
\hspace{-6mm}
\begin{tabular}{cccccc}
\hline
Date   &    UT      & BJD          & \sn\ & $\sigma_{\rm LSD}$ & $r$ \\
(2015) & (hh:mm:ss) & (2,457,340+) &      &   (0.01\%)         &     \\
\hline
Nov 18 & 12:49:11 &  5.03967 & 160 & 2.8 & 0.217 \\
Nov 22 & 10:24:04 &  8.93898 & 180 & 2.4 & 0.902 \\
Nov 23 & 10:24:13 &  9.93909 & 170 & 2.5 & 1.077 \\
Nov 24 & 08:36:48 & 10.86452 & 110 & 4.2 & 1.239 \\
Nov 24 & 09:32:38 & 10.90329 & 150 & 3.1 & 1.246 \\
Nov 25 & 08:35:51 & 11.86388 & 160 & 2.9 & 1.415 \\
Nov 26 & 08:20:02 & 12.85291 & 160 & 2.7 & 1.588 \\
Nov 27 & 07:51:33 & 13.83313 & 160 & 2.8 & 1.760 \\
Nov 28 & 08:46:46 & 14.87149 & 150 & 2.8 & 1.942 \\
Nov 29 & 10:21:57 & 15.93760 & 170 & 2.6 & 2.129 \\
Nov 30 & 11:53:53 & 17.00145 & 160 & 2.8 & 2.316 \\
Dec 01 & 10:23:42 & 17.93882 & 160 & 2.9 & 2.481 \\
Dec 02 & 12:09:32 & 19.01233 & 170 & 2.6 & 2.669 \\
Dec 03 & 09:01:12 & 19.88154 & 180 & 2.3 & 2.821 \\
\hline
\end{tabular}
\label{tab:log}
\end{table}

A few unpolarized (Stokes $I$) spectra (at cycles 1.239, 1.588, 1.760 and 1.942) were affected by 
moonlight in the blue wing of the photospheric lines.  This pollution was filtered with the technique we previously 
devised, which proved efficient and accurate \citep{Donati16, Donati17}.  

\begin{figure}
\includegraphics[scale=0.4,angle=-90]{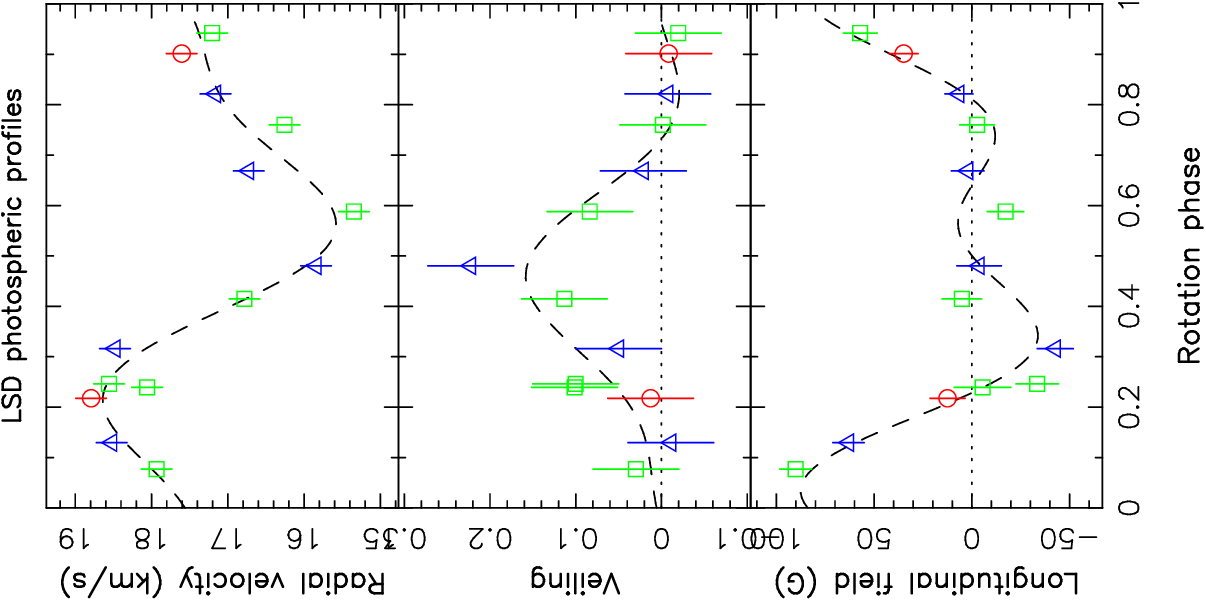}\hspace{3mm}
\includegraphics[scale=0.4,angle=-90]{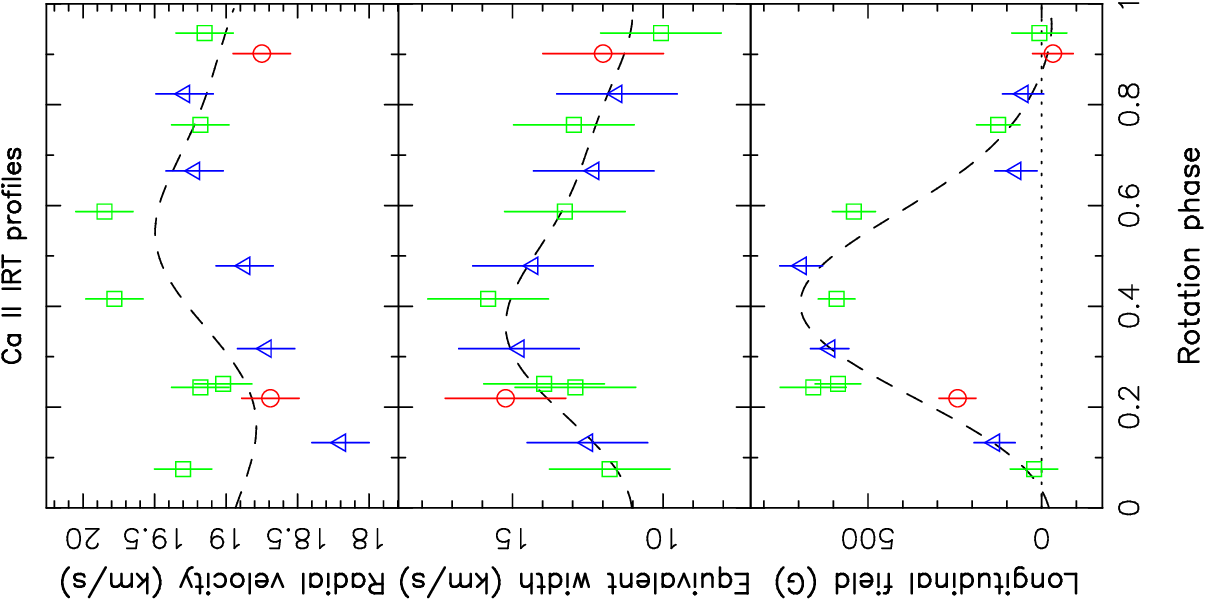} 
\caption[]{Variability of the LSD profiles (left) and \caii\ IRT emission cores (right) of LkCa~15 as a function of 
rotation phase.  Each panel shows the radial velocity (top), the equivalent width (or veiling in the case of LSD profiles, middle) 
and the longitudinal field (bottom).  The red circles, green squares and blue triangles depict measurements obtained during rotation 
cycle 0, 1 and 2 respectively, whereas the dashed line shows a sine plus first harmonic fit to the data.  
Positive longitudinal fields correspond to fields pointing towards the observer.  } 
\label{fig:var}
\end{figure}

Circular polarization (Stokes $V$) LSD profiles all show clear Zeeman signatures indicating the unambiguous 
detection of magnetic fields at the surface of LkCa~15, and revealing longitudinal fields (i.e., line-of-sight 
projected magnetic fields averaged over the visible hemisphere) ranging from $-43$ to 90~G (see Fig.~\ref{fig:var} 
bottom left panel).  Zeeman signals are also clearly detected in the \hei\ $D_3$ emission line (thought to probe 
the footpoints of the magnetic funnels linking the surface of cTTSs to their inner accretion discs) as well as in 
the core emission of the \caii\ IRT lines (presumably probing both the chromosphere and the accretion regions).  
The corresponding longitudinal fields reach up to 0.7~kG and 2~kG for the \caii\ IRT and \hei\ lines respectively at 
phases of maximum emission (0.3-0.5,   
see Fig.~\ref{fig:var} middle and bottom right panel for the \caii\ IRT lines), with the weaker fields in \caii\ IRT 
lines reflecting the fact that emission from the post-shock accretion region is diluted with chromospheric emission 
(whereas \hei\ emission suffers little to no dilution from the chromosphere).  The mass accretion rate 
we derive from the maximum emission fluxes in the \hei\ $D_3$ and \caii\ IRT lines is found to be $10^{-9.2\pm0.3}$~\mspy, 
in good agreement with \citet{Alencar18}.  

Rotational modulation is obvious from LSD photospheric profiles (see Fig.~\ref{fig:var} left panel) and yields rotation periods 
of $5.70\pm0.06$~d, $5.77\pm0.16$~d and $5.63\pm0.12$~d for the longitudinal field, veiling and radial velocity (RV) respectively. 
Modulation is also obvious in \caii\ IRT and \hei\ emission (with periods of $5.82\pm0.16$~d and $5.65\pm0.08$~d 
respectively) and longitudinal fields from \caii\ IRT emission (with period of $5.50\pm0.15$~d).  
Since all periods we derive are quite consistent (within 1.5$\sigma$) with the rotation period used to phase our data (5.70~d, see 
Table~\ref{tab:log}), we report no evidence for differential rotation at the surface of LkCa~15.  

\section{Tomographic modelling of LkCa~15}
\label{sec:tom}

We use tomographic imaging, and more specifically Zeeman-Doppler Imaging (ZDI), to simultaneously reconstruct the large-scale 
magnetic field at the surface of LkCa~15, and the distributions of photospheric brightness and accretion-induced excess 
emission in \caii\ IRT lines.  To achieve this we proceed as in previous studies \citep[e.g.,][]{Donati11, Donati12} where the 
imaging method was outlined.  We thus only recall the main steps in this paper.  

ZDI iteratively looks for a set of images with lowest information content that fits the data at a given reduced chi-square 
level, starting from blank images containing no information.  Whereas photospheric brightness (with cool spots only) and 
accretion-induced excess emission (with bright features only) are directly described through their distributions at the surface 
of the star, the large-scale magnetic field, decomposed into its poloidal and toroidal components, is expressed as a set of 
spherical harmonics \citep[SH,][]{Donati06b}.  We assume the field is dominated by odd SH modes \citep[as in, e.g.,][]{Donati11} 
to ensure that accretion occurs mostly towards the polar regions.  

The local Stokes $I$ and $V$ profiles of both photospheric lines and \caii\ emission are computed
using Unno-Rachkovsky's analytical solution to the polarized radiative transfer equations, taking into account the local 
values of the modeled distributions (i.e., the brightness map for photospheric lines, the accretion map for \caii\ emission, and 
the magnetic map for both sets of lines).  Local profiles are then integrated over the visible hemisphere to obtain the 
synthetic profiles of the rotating star at each observed phase.  

\begin{figure*}
\includegraphics[scale=0.55,angle=-90]{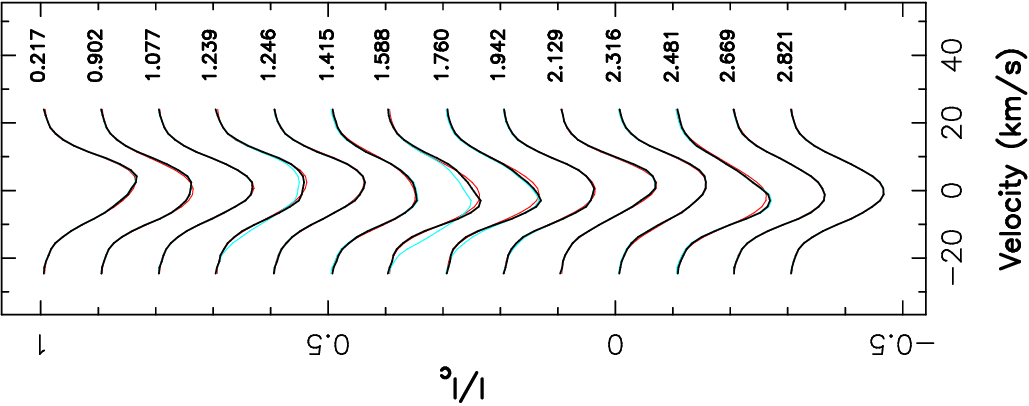}\hspace{2mm}
\includegraphics[scale=0.55,angle=-90]{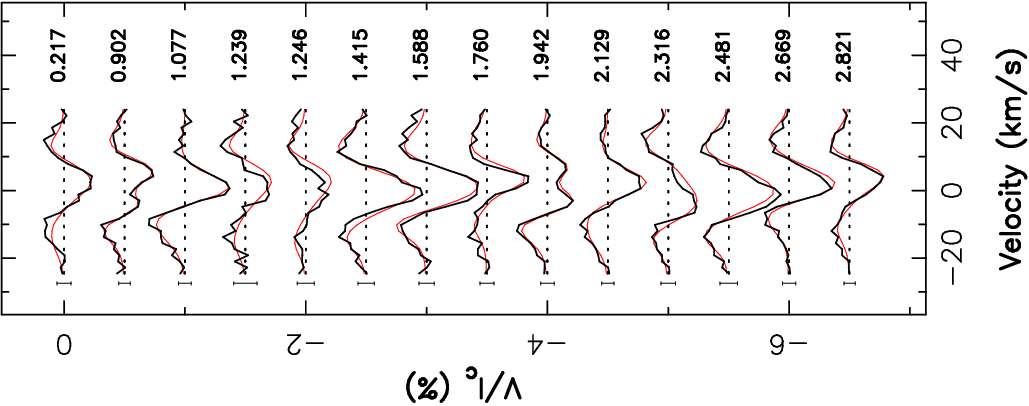}\hspace{5mm}
\includegraphics[scale=0.55,angle=-90]{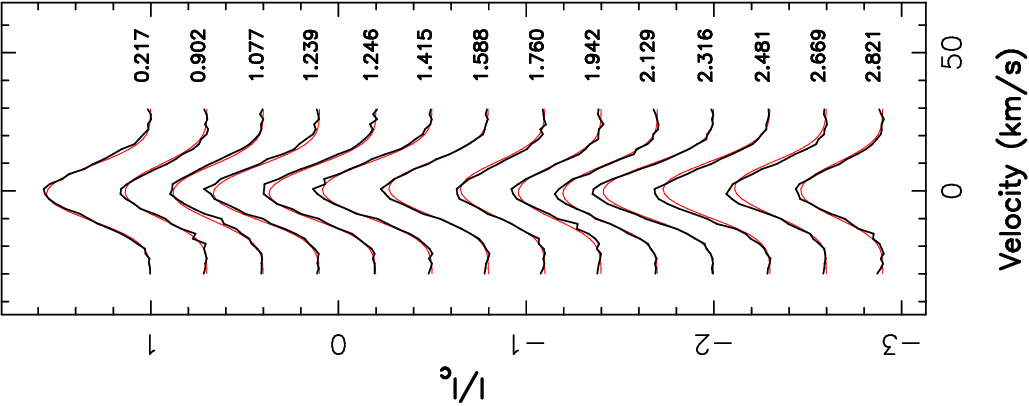}\hspace{2mm}
\includegraphics[scale=0.55,angle=-90]{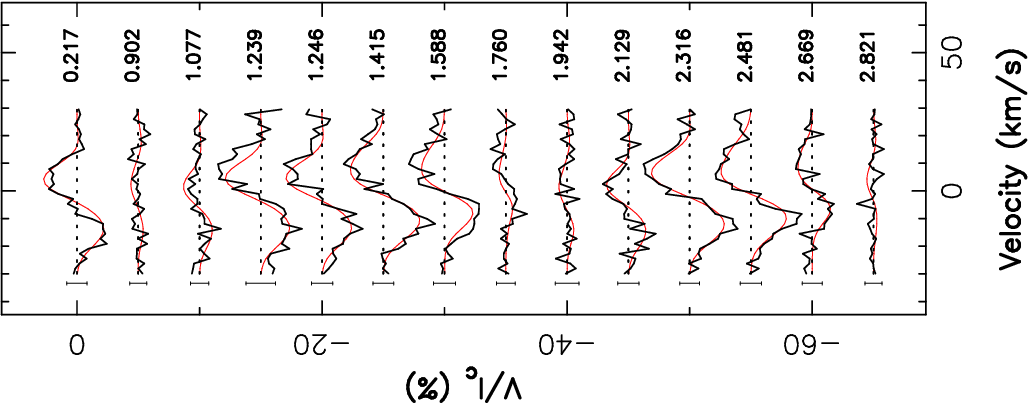}
\caption[]{Observed (thick black line) and modeled (thin red line) LSD Stokes $I$ and $V$ profiles of the photospheric lines 
(left panels) and of the emission core of IRT lines (right panels) of LkCa~15.  Rotation cycles and 3$\sigma$ error bars (for 
Stokes $V$ profiles only) are indicated right and left to each observation respectively.  LSD Stokes $I$ photospheric profiles 
before correcting from moon pollution (at cycles 1.239, 1.588, 1.760 and 1.942) are shown in cyan (leftmost panel). } 
\label{fig:fit}
\end{figure*}

Our model is able to convincingly reproduce most observed profile distortions and Zeeman signatures, as shown in Fig.~\ref{fig:fit}.  
We note that the sharp increase and decrease of the longitudinal field of \caii\ IRT lines are better matched for an inclination angle 
$i$ of 70\degr\ than of 50\degr\ \citep[the angle measured for the outer accretion disc,][with no reliable inclination estimate for  
the inner disk]{Thalmann14}.  We also obtain that the data are 
best fit for $\vsini=12.8\pm0.2$~\kms\ and for an average radial velocity of LkCa~15 with respect to the Sun of $\vrad=17.4\pm0.1$~\kms.  

\begin{figure*}
\includegraphics[scale=0.65,angle=-90,bb=35 40 434 709]{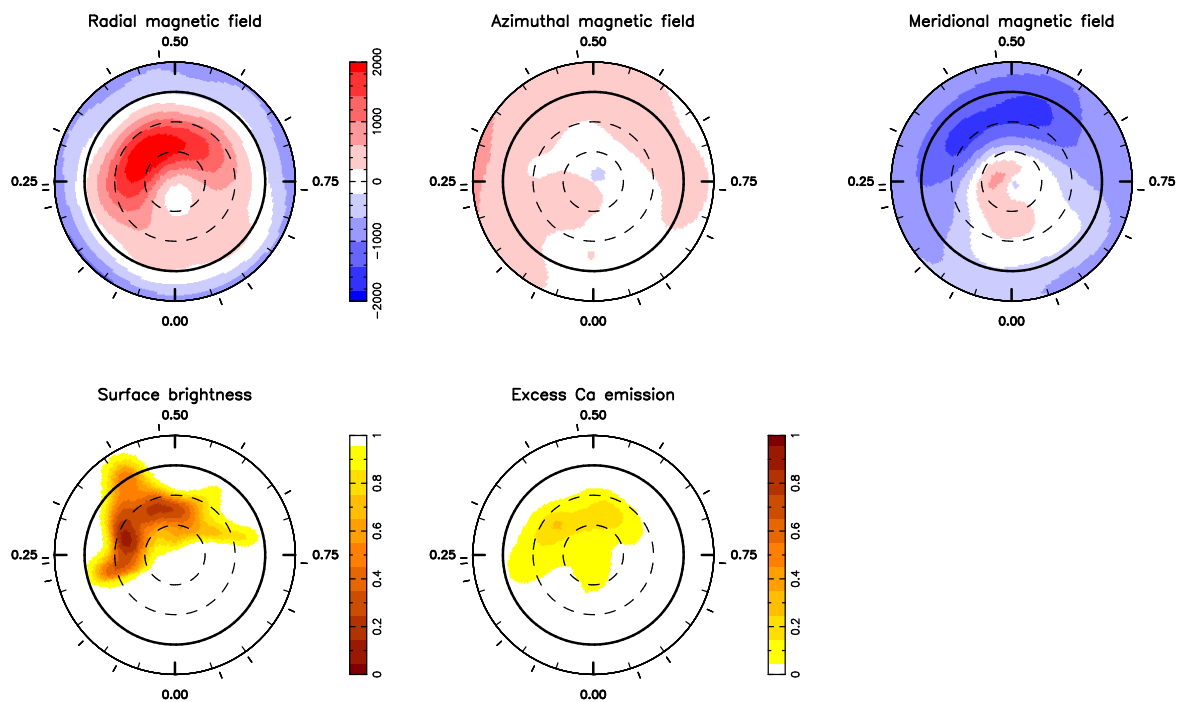}
\caption[]{Reconstructed maps of the magnetic field (top left, middle and right panels for the radial, azimuthal and 
meridional components in spherical coordinates, all in G), relative photospheric brightness (bottom left) and excess 
accretion-induced \caii\ core  emission  (bottom  right)  at  the  surface  of  LkCa~15, derived from the data set of 
Fig.~\ref{fig:fit} using tomographic imaging.  The star is shown in a flattened polar projection down to a latitude of 
$-30$\degr, with the north pole at the center and the equator shown as a bold line.  Outer ticks depict phases of observations.
Positive radial, azimuthal and meridional fields respectively point outwards, counterclockwise and polewards. } 
\label{fig:map}
\end{figure*}

The reconstructed distributions are shown in Fig.~\ref{fig:map}.  The magnetic map shows a strong radial field region at 
intermediate latitudes and phase 0.4 where the field strength reaches 2.2~kG, in agreement with the observed longitudinal 
fields of both \caii\ IRT lines (see Fig.~\ref{fig:var} bottom right panel) and \hei\ line (see Sec.~\ref{sec:obs}).  
This magnetic field region coincides with a large dark photospheric spot, causing longitudinal fields as seen in LSD 
photospheric profiles to be much weaker than those probed by accretion lines;  it also overlaps with a region of 
excess \caii\ emission tracing the footpoints of the magnetic funnel linking the surface of LkCa~15 to the innermost regions 
of the accretion disc.  

We find that the large-scale magnetic field is mostly poloidal, with a poloidal component storing 85\% of the magnetic energy.  
This poloidal component mainly consists of a 1.35~kG dipole tilted at $\simeq$20\degr\ to the rotation axis (towards phase 0.40) 
and  enclosing $\simeq$75\% of the poloidal field energy;  the dipole field strength changes by $\simeq$25\% for a 
10\degr\ change in the assumed inclination.  The poloidal field also includes a $-$0.9~kG octupole component (i.e., anti-parallel 
with the main dipole), gathering $\simeq$20\% of the poloidal field energy, and tilted by $\simeq$20\degr\ to the rotation axis 
(towards phase 0.95).  

\section{Summary and discussion}
\label{sec:dis}

We carried out a spectropolarimetric monitoring of the well-known cTTS LkCa~15 with ESPaDOnS at CFHT to detect and model its 
large-scale magnetic field, and estimate the strength of its dipole component.  From a spectroscopic analysis of our data, we first 
obtain that LkCa~15 is a $\simeq$5~Myr star with a mass of $1.25\pm0.10$~\msun\ and a radius of $1.6\pm0.2$~\rsun\ according to the 
evolution models of \citet{Siess00}, and thus appears as a slightly older and less massive version of V2129~Oph, another well-studied 
cTTS \citep{Donati11}.  

Zeeman signatures from LkCa~15 are clearly detected, both in photospheric and accretion lines.  Our tomographic study reveals that 
LkCa~15 hosts a strong and mostly poloidal large-scale field, reaching up to 2.2~kG at the surface of the star, with a dipole component 
of 1.35~kG.  Given the accretion rate at the surface of the star, found to be equal to $10^{-9.2}$~\mspy\ from emission 
fluxes in the \hei\ and \caii\ IRT lines (see Sec.~\ref{sec:obs}), we can conclude from \citet{Bessolaz08} that the magnetospheric 
cavity that the field carves at the disc center reaches out to the corotation radius, located at a distance of 0.07~au.  Our result 
suggests that LkCa~15 is in a propeller accretion regime, capable of counteracting the natural angular acceleration of the 
contracting star.  In fact, LkCa~15 is a good match to the C01 simulation of \citet{Zanni13} where the spin-down torque from the star-disc 
coupling is found to dominate the spin-up torque from both the accreted material and the contracting star.  Besides, photometric 
monitoring tells that accretion at the surface of LkCa~15 is unsteady \citep[][as for AA~Tau]{Alencar18}, agreeing again 
with predictions from simulations \citep{Zanni13}.  This is also consistent with recent studies concluding that LkCa~15 triggers 
outflows from the inner disc \citep{Fang18, Mendigutia18}.  

Given that LkCa~15 is already close to 50\% convective in radius (see Sec.~\ref{sec:evo}), we can speculate that the dipole component 
of its large-scale field is likely to be strongly reduced as the star contracts and becomes mostly radiative \citep{Donati09, Gregory12, 
Donati13}.  As a result, LkCa~15 is likely to start speeding up towards the main sequence in a few Myr.  
Simulations are needed to investigate in a more documented and quantitative way the rotational history of cTTSs like LkCa~15 under 
the combined effect of irregular accretion and evolving large-scale fields.  

We stress the importance of studying the large-scale fields of a wide sample of cTTSs like LkCa~15 to better 
understand the physics of star-disc interactions, and more generally its impact on early stellar evolution.  Studying 
how magnetic fields relate to the various types of magnetospheric accretion revealed through continuous photometry \citep{Cody14, 
Sousa16, Cody18} seems particularly promising in this respect.  New generation spectropolarimeters working in the nIR like SPIRou 
\citep{Donati17b}, giving access to even younger and lower-mass stars and offering an enhanced sensitivity to magnetic fields, 
should be a prime asset for this task.  

\section*{Acknowledgements}
We thank the referee for remarks that clarified the paper.  
Our study is based on data obtained at the CFHT, operated by the CNRC (Canada), INSU/CNRS (France) and the University of Hawaii.  
This project received funding from the European Research Council (ERC) under the H2020 research \& innovation programme (grant agreements 
\#740651 NewWorlds and \#742095 SPIDI).  SHPA acknowledges financial support from CNPq, CAPES and Fapemig.  
We also thank the Programme National de Physique Stellaire (PNPS) of CNRS/INSU for financial support.  

\bibliography{lkca15}

\begin{thebibliography}{}
\makeatletter
\relax
\def\mn@urlcharsother{\let\do\@makeother \do\$\do\&\do\#\do\^\do\_\do\%\do\~}
\def\mn@doi{\begingroup\mn@urlcharsother \@ifnextchar [ {\mn@doi@}
  {\mn@doi@[]}}
\def\mn@doi@[#1]#2{\def\@tempa{#1}\ifx\@tempa\@empty \href
  {http://dx.doi.org/#2} {doi:#2}\else \href {http://dx.doi.org/#2} {#1}\fi
  \endgroup}
\def\mn@eprint#1#2{\mn@eprint@#1:#2::\@nil}
\def\mn@eprint@arXiv#1{\href {http://arxiv.org/abs/#1} {{\tt arXiv:#1}}}
\def\mn@eprint@dblp#1{\href {http://dblp.uni-trier.de/rec/bibtex/#1.xml}
  {dblp:#1}}
\def\mn@eprint@#1:#2:#3:#4\@nil{\def\@tempa {#1}\def\@tempb {#2}\def\@tempc
  {#3}\ifx \@tempc \@empty \let \@tempc \@tempb \let \@tempb \@tempa \fi \ifx
  \@tempb \@empty \def\@tempb {arXiv}\fi \@ifundefined
  {mn@eprint@\@tempb}{\@tempb:\@tempc}{\expandafter \expandafter \csname
  mn@eprint@\@tempb\endcsname \expandafter{\@tempc}}}

\bibitem[\protect\citeauthoryear{{Alencar}, {Bouvier}, {Alecian}, {Donati},
  {Folsom}, {Grankin}  \& {et al.,}}{{Alencar} et~al.}{2018}]{Alencar18}
{Alencar} S.,  {Bouvier} J.,  {Alecian} E.,  {Donati} J.-F.,  {Folsom} C.,
  {Grankin} K.,   {et al.,} 2018, \aap, p. submitted

\bibitem[\protect\citeauthoryear{{Bessolaz}, {Zanni}, {Ferreira}, {Keppens}  \&
  {Bouvier}}{{Bessolaz} et~al.}{2008}]{Bessolaz08}
{Bessolaz} N.,  {Zanni} C.,  {Ferreira} J.,  {Keppens} R.,   {Bouvier} J.,
  2008, \mn@doi [\aap] {10.1051/0004-6361:20078328}, \href
  {http://adsabs.harvard.edu/abs/2008A%26A...478..155B} {478, 155}

\bibitem[\protect\citeauthoryear{{Bouvier} et~al.,}{{Bouvier}
  et~al.}{2007}]{Bouvier07b}
{Bouvier} J.,  et~al., 2007, \mn@doi [\aap] {10.1051/0004-6361:20066021}, \href
  {http://adsabs.harvard.edu/abs/2007A%26A...463.1017B} {463, 1017}

\bibitem[\protect\citeauthoryear{{Bouvier}, {Matt}, {Mohanty}, {Scholz},
  {Stassun}  \& {Zanni}}{{Bouvier} et~al.}{2014}]{Bouvier14}
{Bouvier} J.,  {Matt} S.~P.,  {Mohanty} S.,  {Scholz} A.,  {Stassun} K.~G.,
  {Zanni} C.,  2014, \mn@doi [Protostars and Planets VI]
  {10.2458/azu_uapress_9780816531240-ch019}, \href
  {http://adsabs.harvard.edu/abs/2014prpl.conf..433B} {pp 433--450}

\bibitem[\protect\citeauthoryear{{Cody} \& {Hillenbrand}}{{Cody} \&
  {Hillenbrand}}{2018}]{Cody18}
{Cody} A.~M.,  {Hillenbrand} L.~A.,  2018, \mn@doi [\aj]
  {10.3847/1538-3881/aacead}, \href
  {http://adsabs.harvard.edu/abs/2018AJ....156...71C} {156, 71}

\bibitem[\protect\citeauthoryear{{Cody} et~al.,}{{Cody} et~al.}{2014}]{Cody14}
{Cody} A.~M.,  et~al., 2014, \mn@doi [\aj] {10.1088/0004-6256/147/4/82}, \href
  {http://adsabs.harvard.edu/abs/2014AJ....147...82C} {147, 82}

\bibitem[\protect\citeauthoryear{{Donati}}{{Donati}}{2003}]{Donati03}
{Donati} J.-F.,  2003, in {Trujillo-Bueno} J.,  {Sanchez Almeida} J.,  eds,
  Astronomical Society of the Pacific Conference Series Vol. 307, Astronomical
  Society of the Pacific Conference Series. p.~41

\bibitem[\protect\citeauthoryear{{Donati} \& {Landstreet}}{{Donati} \&
  {Landstreet}}{2009}]{Donati09}
{Donati} J.,  {Landstreet} J.~D.,  2009, \mn@doi [\araa]
  {10.1146/annurev-astro-082708-101833}, \href
  {http://adsabs.harvard.edu/abs/2009ARA%26A..47..333D} {47, 333}

\bibitem[\protect\citeauthoryear{{Donati}, {Semel}, {Carter}, {Rees}  \&
  {Collier Cameron}}{{Donati} et~al.}{1997}]{Donati97b}
{Donati} J.-F.,  {Semel} M.,  {Carter} B.~D.,  {Rees} D.~E.,   {Collier
  Cameron} A.,  1997, \mnras, \href
  {http://adsabs.harvard.edu/abs/1997MNRAS.291..658D} {291, 658}

\bibitem[\protect\citeauthoryear{{Donati} et~al.,}{{Donati}
  et~al.}{2006}]{Donati06b}
{Donati} J.-F.,  et~al., 2006, \mn@doi [\mnras]
  {10.1111/j.1365-2966.2006.10558.x}, \href
  {http://adsabs.harvard.edu/abs/2006MNRAS.370..629D} {370, 629}

\bibitem[\protect\citeauthoryear{{Donati} et~al.,}{{Donati}
  et~al.}{2010}]{Donati10}
{Donati} J.,  et~al., 2010, \mn@doi [\mnras]
  {10.1111/j.1365-2966.2009.15998.x}, \href
  {http://adsabs.harvard.edu/abs/2010MNRAS.402.1426D} {402, 1426}

\bibitem[\protect\citeauthoryear{{Donati} et~al.,}{{Donati}
  et~al.}{2011}]{Donati11}
{Donati} J.,  et~al., 2011, \mn@doi [\mnras]
  {10.1111/j.1365-2966.2010.18069.x}, \href
  {http://adsabs.harvard.edu/abs/2011MNRAS.412.2454D} {412, 2454}

\bibitem[\protect\citeauthoryear{{Donati} et~al.,}{{Donati}
  et~al.}{2012}]{Donati12}
{Donati} J.-F.,  et~al., 2012, \mn@doi [\mnras]
  {10.1111/j.1365-2966.2012.21482.x}, \href
  {http://adsabs.harvard.edu/abs/2012MNRAS.425.2948D} {425, 2948}

\bibitem[\protect\citeauthoryear{{Donati} et~al.,}{{Donati}
  et~al.}{2013}]{Donati13}
{Donati} J.-F.,  et~al., 2013, \mn@doi [\mnras] {10.1093/mnras/stt1622}, \href
  {http://adsabs.harvard.edu/abs/2013MNRAS.436..881D} {436, 881}

\bibitem[\protect\citeauthoryear{{Donati} et~al.,}{{Donati}
  et~al.}{2014}]{Donati14}
{Donati} J.-F.,  et~al., 2014, \mn@doi [\mnras] {10.1093/mnras/stu1679}, \href
  {http://adsabs.harvard.edu/abs/2014MNRAS.444.3220D} {444, 3220}

\bibitem[\protect\citeauthoryear{{Donati} et~al.,}{{Donati}
  et~al.}{2016}]{Donati16}
{Donati} J.~F.,  et~al., 2016, \mn@doi [\nat] {10.1038/nature18305}, \href
  {http://adsabs.harvard.edu/abs/2016Natur.534..662D} {534, 662}

\bibitem[\protect\citeauthoryear{{Donati}, {Kouach}, {Lacombe}, {Baratchart},
  {Doyon}, {Delfosse}  \& {et al.}}{{Donati} et~al.}{2017a}]{Donati17b}
{Donati} J.-F.,  {Kouach} D.,  {Lacombe} M.,  {Baratchart} S.,  {Doyon} R.,
  {Delfosse} X.,   {et al.} 2017a, {SPIRou: A nIR
  Spectropolarimeter/High-precision Velocimeter for the CFHT}.
p.~107, \mn@doi{10.1007/978-3-319-30648-3_107-1}

\bibitem[\protect\citeauthoryear{{Donati} et~al.,}{{Donati}
  et~al.}{2017b}]{Donati17}
{Donati} J.-F.,  et~al., 2017b, \mn@doi [\mnras] {10.1093/mnras/stw2904}, \href
  {http://adsabs.harvard.edu/abs/2017MNRAS.465.3343D} {465, 3343}

\bibitem[\protect\citeauthoryear{{Esau}, {Harries}  \& {Bouvier}}{{Esau}
  et~al.}{2014}]{Esau14}
{Esau} C.~F.,  {Harries} T.~J.,   {Bouvier} J.,  2014, \mn@doi [\mnras]
  {10.1093/mnras/stu1211}, \href
  {http://adsabs.harvard.edu/abs/2014MNRAS.443.1022E} {443, 1022}

\bibitem[\protect\citeauthoryear{{Fang} et~al.,}{{Fang} et~al.}{2018}]{Fang18}
{Fang} M.,  et~al., 2018, preprint, \href
  {http://adsabs.harvard.edu/abs/2018arXiv181003366F} {} (\mn@eprint {arXiv}
  {1810.03366})

\bibitem[\protect\citeauthoryear{{Gaia Collaboration}}{{Gaia
  Collaboration}}{2018}]{Gaia18}
{Gaia Collaboration} 2018, VizieR Online Data Catalog, \href
  {http://adsabs.harvard.edu/abs/2018yCat.1345....0G} {1345}

\bibitem[\protect\citeauthoryear{{Grankin}, {Bouvier}, {Herbst}  \&
  {Melnikov}}{{Grankin} et~al.}{2008}]{Grankin08}
{Grankin} K.~N.,  {Bouvier} J.,  {Herbst} W.,   {Melnikov} S.~Y.,  2008,
  \mn@doi [\aap] {10.1051/0004-6361:20078476}, \href
  {http://adsabs.harvard.edu/abs/2008A%26A...479..827G} {479, 827}

\bibitem[\protect\citeauthoryear{{Gregory}, {Donati}, {Morin}, {Hussain},
  {Mayne}, {Hillenbrand}  \& {Jardine}}{{Gregory} et~al.}{2012}]{Gregory12}
{Gregory} S.~G.,  {Donati} J.-F.,  {Morin} J.,  {Hussain} G.~A.~J.,  {Mayne}
  N.~J.,  {Hillenbrand} L.~A.,   {Jardine} M.,  2012, \mn@doi [\apj]
  {10.1088/0004-637X/755/2/97}, \href
  {http://adsabs.harvard.edu/abs/2012ApJ...755...97G} {755, 97}

\bibitem[\protect\citeauthoryear{{Kraus} \& {Ireland}}{{Kraus} \&
  {Ireland}}{2012}]{Kraus12b}
{Kraus} A.~L.,  {Ireland} M.~J.,  2012, \mn@doi [\apj]
  {10.1088/0004-637X/745/1/5}, \href
  {http://adsabs.harvard.edu/abs/2012ApJ...745....5K} {745, 5}

\bibitem[\protect\citeauthoryear{{Mendigut{\'{\i}}a}, {Oudmaijer}, {Schneider},
  {Hu{\'e}lamo}, {Baines}, {Brittain}  \& {Aberasturi}}{{Mendigut{\'{\i}}a}
  et~al.}{2018}]{Mendigutia18}
{Mendigut{\'{\i}}a} I.,  {Oudmaijer} R.~D.,  {Schneider} P.~C.,  {Hu{\'e}lamo}
  N.,  {Baines} D.,  {Brittain} S.~D.,   {Aberasturi} M.,  2018, preprint,
  \href {http://adsabs.harvard.edu/abs/2018arXiv181004181M} {} (\mn@eprint
  {arXiv} {1810.04181})

\bibitem[\protect\citeauthoryear{{Oh} et~al.,}{{Oh} et~al.}{2016}]{Oh16}
{Oh} D.,  et~al., 2016, \mn@doi [\pasj] {10.1093/pasj/psv133}, \href
  {http://adsabs.harvard.edu/abs/2016PASJ...68L...3O} {68, L3}

\bibitem[\protect\citeauthoryear{{Pecaut} \& {Mamajek}}{{Pecaut} \&
  {Mamajek}}{2013}]{Pecaut13}
{Pecaut} M.~J.,  {Mamajek} E.~E.,  2013, \mn@doi [\apjs]
  {10.1088/0067-0049/208/1/9}, \href
  {http://adsabs.harvard.edu/abs/2013ApJS..208....9P} {208, 9}

\bibitem[\protect\citeauthoryear{{Romanova}, {Ustyugova}, {Koldoba}  \&
  {Lovelace}}{{Romanova} et~al.}{2004}]{Romanova04}
{Romanova} M.~M.,  {Ustyugova} G.~V.,  {Koldoba} A.~V.,   {Lovelace} R.~V.~E.,
  2004, \mn@doi [\apjl] {10.1086/426586}, \href
  {http://adsabs.harvard.edu/abs/2004ApJ...616L.151R} {616, L151}

\bibitem[\protect\citeauthoryear{{Sallum} et~al.,}{{Sallum}
  et~al.}{2015}]{Sallum15}
{Sallum} S.,  et~al., 2015, \mn@doi [\nat] {10.1038/nature15761}, \href
  {http://adsabs.harvard.edu/abs/2015Natur.527..342S} {527, 342}

\bibitem[\protect\citeauthoryear{{Siess}, {Dufour}  \& {Forestini}}{{Siess}
  et~al.}{2000}]{Siess00}
{Siess} L.,  {Dufour} E.,   {Forestini} M.,  2000, \aap, 358, 593

\bibitem[\protect\citeauthoryear{{Sousa} et~al.,}{{Sousa}
  et~al.}{2016}]{Sousa16}
{Sousa} A.~P.,  et~al., 2016, \mn@doi [\aap] {10.1051/0004-6361/201526599},
  \href {http://adsabs.harvard.edu/abs/2016A%26A...586A..47S} {586, A47}

\bibitem[\protect\citeauthoryear{{Thalmann} et~al.,}{{Thalmann}
  et~al.}{2014}]{Thalmann14}
{Thalmann} C.,  et~al., 2014, \mn@doi [\aap] {10.1051/0004-6361/201322915},
  \href {http://adsabs.harvard.edu/abs/2014A%26A...566A..51T} {566, A51}

\bibitem[\protect\citeauthoryear{{Thalmann} et~al.,}{{Thalmann}
  et~al.}{2015}]{Thalmann15}
{Thalmann} C.,  et~al., 2015, \mn@doi [\apjl] {10.1088/2041-8205/808/2/L41},
  \href {http://adsabs.harvard.edu/abs/2015ApJ...808L..41T} {808, L41}

\bibitem[\protect\citeauthoryear{{Thalmann} et~al.,}{{Thalmann}
  et~al.}{2016}]{Thalmann16}
{Thalmann} C.,  et~al., 2016, \mn@doi [\apjl] {10.3847/2041-8205/828/2/L17},
  \href {http://adsabs.harvard.edu/abs/2016ApJ...828L..17T} {828, L17}

\bibitem[\protect\citeauthoryear{{Ustyugova}, {Koldoba}, {Romanova}  \&
  {Lovelace}}{{Ustyugova} et~al.}{2006}]{Ustyugova06}
{Ustyugova} G.~V.,  {Koldoba} A.~V.,  {Romanova} M.~M.,   {Lovelace} R.~V.~E.,
  2006, \mn@doi [\apj] {10.1086/503379}, \href
  {http://adsabs.harvard.edu/abs/2006ApJ...646..304U} {646, 304}

\bibitem[\protect\citeauthoryear{{Zanni} \& {Ferreira}}{{Zanni} \&
  {Ferreira}}{2013}]{Zanni13}
{Zanni} C.,  {Ferreira} J.,  2013, \mn@doi [\aap]
  {10.1051/0004-6361/201220168}, \href
  {http://adsabs.harvard.edu/abs/2013A%26A...550A..99Z} {550, A99}

\bibitem[\protect\citeauthoryear{{van der Marel}, {van Dishoeck}, {Bruderer},
  {P{\'e}rez}  \& {Isella}}{{van der Marel} et~al.}{2015}]{vanderMarel15}
{van der Marel} N.,  {van Dishoeck} E.~F.,  {Bruderer} S.,  {P{\'e}rez} L.,
  {Isella} A.,  2015, \mn@doi [\aap] {10.1051/0004-6361/201525658}, \href
  {http://adsabs.harvard.edu/abs/2015A%26A...579A.106V} {579, A106}

\makeatother
\end{thebibliography}
\bibliographystyle{mnras}

\bsp	
\label{lastpage}
\end{document}